\pdfoutput=1
\documentclass[PAPER,cernpreprint, atlasdraft=false, UKenglish, texmf, orcidlogo]{atlasdoc}
\usepackage{atlaspackage}
\usepackage{rotating}
\usepackage{multirow}
\usepackage{atlasbiblatex}

\usepackage{atlasphysics}
\graphicspath{{logos/}{figures/}}

\usepackage{ANA-EXOT-2019-02-PAPER-defs}

\AtlasTitle{Search for pair-production of vector-like $T$ quarks decaying into a top quark and a spin-0 particle in the diphoton final state in proton proton collisions at $\sqrt{s}$=13 TeV with the
ATLAS detector}

\AtlasAbstract{%
A search for pair-produced vector-like T quarks (VLQs) is presented. Each VLQ is assumed to decay into a top quark and an exotic (pseudo)scalar S. The search targets events with at least one resonant $S \to \gamma\gamma$ decay and large summed transverse momentum and missing transverse momentum from other final-state particles.
The analysis uses data from $pp$ collisions at a centre-of-mass energy of $\sqrt{s}$ = 13 TeV recorded by the ATLAS detector from 2015 to 2018 and corresponding to an integrated luminosity of 140 fb$^{-1}$.
This is the first VLQ search performed in this final state. No significant excess over the Standard Model background is observed.
The analysis is model-independent to a large extent but targets cascade decays into
final states with electroweak bosons or $t\bar{t}$.
Results are provided in the form of upper limits on the fiducial cross-section times branching ratio as a function of the masses of the VLQ and the spin-0 particle. The fiducial volume is defined by kinematic requirements on the photons from the spin-0 decay.
}

\AtlasRefCode{EXOT-2019-02}

\PreprintIdNumber{CERN-EP-2026-225}

\AtlasJournal{JHEP}
\AtlasCoverEgroupAnalysisTeam{atlas-exot-2019-02-analysis-team@cern.ch}

\hypersetup{pdftitle={ATLAS document},pdfauthor={The ATLAS Collaboration}}

\begin{document}

\maketitle

\tableofcontents

\section{Introduction}
%

%

After the discovery of the Higgs boson by the ATLAS and CMS experiments in 2012~\cite{HIGG-2012-27,CMS-HIG-12-028}, exploring its nature has become a central theme for high-energy physics.
An obvious line of inquiry is whether the Higgs boson is a fundamental or composite particle. Another mystery is the low mass of the Higgs boson leading to the so-called hierarchy problem or the fine-tuning problem. Popular proposed solutions include supersymmetry and compositeness ~\cite{Kaplan:1983fs}.

Furthermore, the very large mass of the top quark relative to all the other fermions is another aspect of the Standard Model (SM) that calls for explanation.
Considering that fermion masses are related to the Higgs sector through the Yukawa terms, theories of new physics beyond the SM very often address both the Higgs and the top sectors at the same time.
Examples include partial compositeness scenarios, where the top quark mixes with fermion condensates in the strong sector of composite Higgs scenarios~\cite{Kaplan:1991dc,Agashe:2004rs}.
A notable prediction of these models is %
the presence of Vector-like quarks (VLQs), also referred to as %
top partners when they have electric charge $2/3$. If VLQs exist, they could be produced in high-energy particle collisions at the Large Hadron Collider (LHC) ~\cite{Evans:2008zzb}, and would then decay into a SM top quark together with a gauge boson or a Higgs boson.

The ATLAS and CMS experiments have conducted many searches for pair-produced VLQs ~\cite{CMS-B2G-19-005,EXOT-2018-58,CMS-B2G-20-011,EXOT-2019-06,EXOT-2019-08,EXOT-2014-10,EXOT-2021-35,EXOT-2017-14,CMS-B2G-18-005,EXOT-2016-14,EXOT-2016-16,CMS-B2G-17-011,CMS-B2G-17-012,EXOT-2016-15,EXOT-2016-13,CMS-B2G-20-014,EXOT-2017-34,CMS-B2G-17-014,CMS-B2G-17-003,EXOT-2016-35}
and singly-produced VLQs \cite{EXOT-2016-35,EXOT-2019-07,EXOT-2020-01,EXOT-2018-52,EXOT-2022-40,CMS-B2G-19-004,CMS-B2G-21-007,CMS-B2G-18-003,CMS-B2G-17-007,EXOT-2017-16,CMS-B2G-19-001,EXOT-2016-17,CMS-B2G-16-006,EXOT-2019-04,CMS-B2G-17-009,CMS-B2G-17-018,EXOT-2018-60,EXOT-2022-43,EXOT-2022-40,EXOT-2021-02,CMS-B2G-23-009,CMS-B2G-22-001}.
All these analyses, except Refs.~\cite{CMS-B2G-23-009,CMS-B2G-22-001} %
assume decays of the VLQs into SM particles, typically a third-generation quark ($t$ or $b$) and a heavy boson ($H$, $W$ or $Z$).
However, if the VLQs arise as a consequence of compositeness, additional exotic scalars or pseudoscalars $S$ may also exist~\cite{Gripaios:2009pe,Barnard:2013zea,Ferretti:2013kya,Serra:2015xfa,Anandakrishnan:2015yfa,Banerjee:2016wls,Dobrescu:2016pda,Chala:2017xgc,Colucci:2018vxz,Banerjee:2018fsx,Han:2018hcu,Kim:2019oyh,Bizot:2018tds,Benbrik:2019zdp,Banerjee:2022izw,Banerjee:2023upj,Banerjee:2023ipb}.
A top partner %
VLQ $T$ could decay into a top quark and new spin-0 particle$S$, $T \rightarrow St$, where the $S$ decays into SM particles.
If these decays are present in nature, the limits from previous searches, including those from the ATLAS VLQ combination, may not apply~\cite{Aguilar-Saavedra:2017giu,Cacciapaglia:2019zmj,Banerjee:2024zvg}.

\begin{figure}[htbp]
\begin{center}
\includegraphics[width=0.55\textwidth]{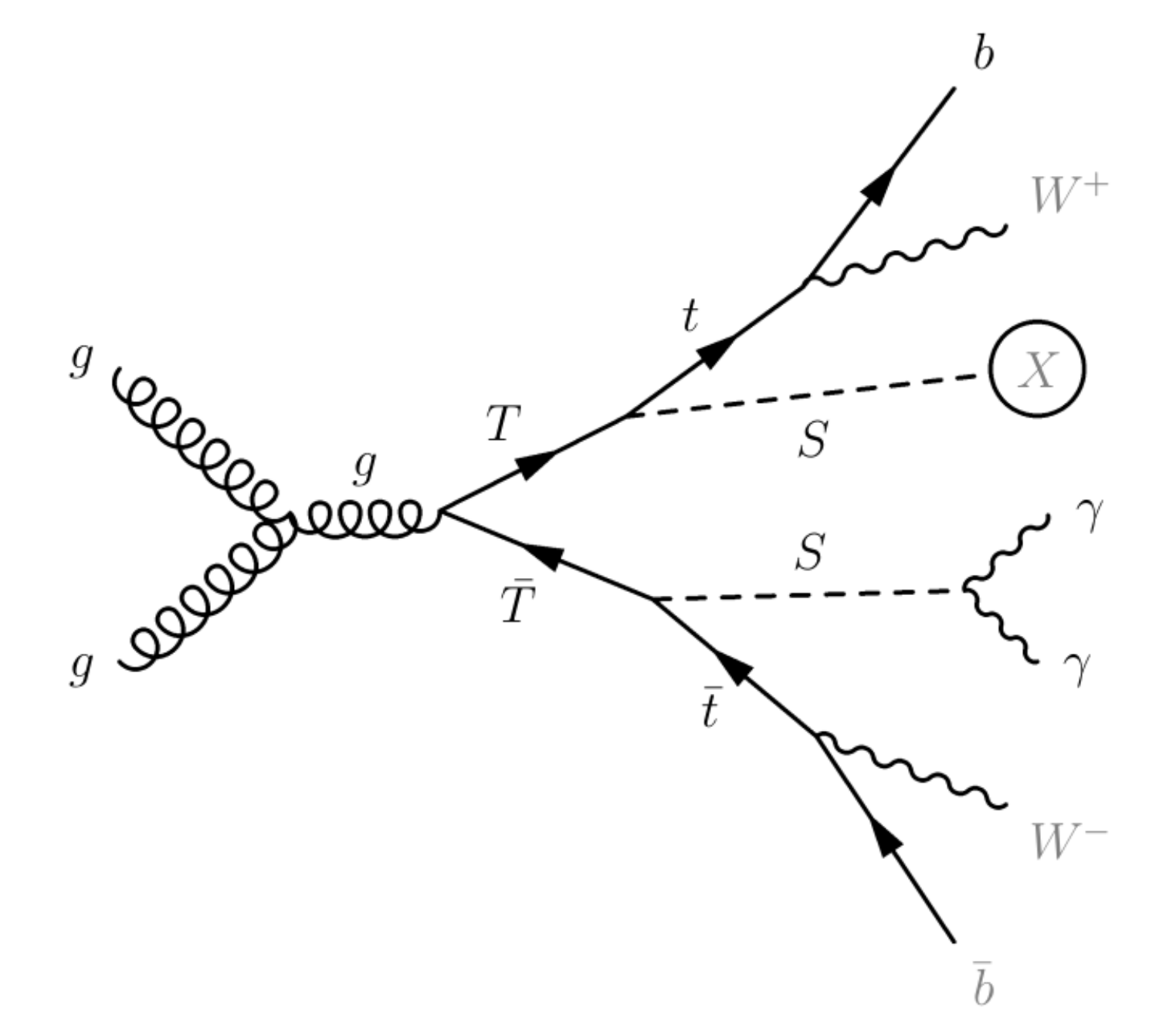}
\end{center}
\caption{Feynman diagram for the top partner VLQ $T\overline{T}$ pair-production and their subsequent decays. One of the $T$s decays into a top quark and a new exotic spin-0 particle $S$, which in turn decays into two photons. The other $T$ follows the same decay chain, the $S$ decays into any of its allowed decay channels denoted by $X$ (\gamgam, $\Zgam$, $WW$, $ZZ$ or $\ttbar$).
\label{fig:PairProdTTgammagammaSInclusive} }
\end{figure}

In the analysis described here, the full Run 2 dataset collected by the ATLAS experiment~\cite{PERF-2007-01} is used to search for pair-production of VLQ $T$  process as illustrated in Figure~\ref{fig:PairProdTTgammagammaSInclusive}, where one $T$ decays via $T \rightarrow St$, $S \rightarrow \gamgam$ and the other  decays via $T \rightarrow St$, $S \to X$, where $X$ is any allowed final state (\gamgam, $\Zgam$, $WW$, $ZZ$ or $\ttbar$). Decays into other SM fermions are negligible, since they are suppressed by the fermion masses.
No distinction is made between the decay of $T$ and $\bar{T}$ throughout the paper.

The final state thus contains two high-energy photons from the $S$ decay, %
two high-energy $b$-quarks and high transverse energy from the other decay products. There may also be substantial missing transverse momentum (\MET) from
neutrinos escaping the detector from leptonic decays of $W$ bosons or $Z\to\nu\nu$ decays.
In this analysis %
$T$ is assumed to decay into $St$ with a 100\% branching ratio.
The main search strategy is to look for an excess over the diphoton invariant mass spectrum in events with large transverse energy and \MET.

The ranges considered for the VLQ $T$ and $S$ masses, $m_T$ and $m_S$, span $m_T \in [1000, 1800]~\GeV$ and $m_S \in [200, 1200]~\GeV$, respectively.
Regions that are kinematically suppressed, such as $m_S > m_T - m_t$, are excluded.
This choice is primarily motivated by phenomenological studies in Ref.~\cite{Benbrik:2019zdp}, which consider the same theoretical framework and indicate that the expected sensitivity is largely concentrated within this region of parameter space.
For a more model-independent interpretation, results are presented in a fiducial volume in which the two leading photons are required to arise from the decay of the same spin-0 particle.

This paper is organised as follows.  Section~\ref{sec:detector} gives an overview of the ATLAS detector, while Section~\ref{sec:data_mc} describes the data set and the simulated samples used. The event reconstruction and object selection are described in Section~\ref{sec:event_object}. The modelling of the signal and the background is described in Section~\ref{sec:signal_model}. %
The treatment of the systematic uncertainties and the statistical procedure are being discussed in Section~\ref{sec:stats}.
The results are presented in Section~\ref{sec:results} and finally the conclusions are reported in Section~\ref{sec:conclusion}.


%
\section{ATLAS detector}
\label{sec:detector}
%

%
%
%
\newcommand{\AtlasCoordFootnote}{%
ATLAS uses a right-handed coordinate system with its origin at the nominal interaction point (IP)
in the centre of the detector and the \(z\)-axis along the beam pipe.
The \(x\)-axis points from the IP to the centre of the LHC ring,
and the \(y\)-axis points upwards.
Polar coordinates \((r,\phi)\) are used in the transverse plane,
\(\phi\) being the azimuthal angle around the \(z\)-axis
and $r$ the distance from the IP in the transverse plane.
The pseudorapidity is defined in terms of the polar angle \(\theta\) as \(\eta = -\ln \tan(\theta/2)\) and is equal to the rapidity
$ y = \frac{1}{2} \ln \left( \frac{E + p_z}{E - p_z} \right) $ in the relativistic limit.
Angular distance is measured in units of \(\Delta R \equiv \sqrt{(\Delta y)^{2} + (\Delta\phi)^{2}}\).}

The ATLAS detector~\cite{PERF-2007-01} at the LHC covers nearly the entire solid angle around the collision point.\footnote{\AtlasCoordFootnote}
It consists of an inner tracking detector surrounded by a thin superconducting solenoid, electromagnetic and hadronic calorimeters,
and a muon spectrometer incorporating three large superconducting air-core toroidal magnets.

The inner-detector system (ID) is immersed in a \qty{2}{\tesla} axial magnetic field
and provides charged-particle tracking in the range of \(|\eta| < 2.5\).
The high-granularity silicon pixel detector covers the vertex region and typically provides four measurements per track,
the first hit generally being in the insertable B-layer (IBL) installed before Run~2~\cite{ATLAS-TDR-19,PIX-2018-001}.
It is followed by the SemiConductor Tracker (SCT), which usually provides eight measurements per track.
These silicon detectors are complemented by the transition radiation tracker (TRT),
which enables radially extended track reconstruction up to \(|\eta| = 2.0\).
The TRT also provides electron identification information
based on the fraction of hits (typically 30 in total) above a higher energy-deposit threshold corresponding to transition radiation.

The calorimeter system covers the pseudorapidity range \(|\eta| < 4.9\).
Within the region \(|\eta|< 3.2\), electromagnetic calorimetry is provided by barrel and
endcap high-granularity lead/liquid-argon (LAr) calorimeters,
with an additional thin LAr presampler covering \(|\eta| < 1.8\)
to correct for energy loss in material upstream of the calorimeters.
Hadronic calorimetry is provided by the steel/scintillator-tile calorimeter,
segmented into three barrel structures within \(|\eta| < 1.7\), and two copper/LAr hadronic endcap calorimeters.
The solid angle coverage is completed with forward copper/LAr and tungsten/LAr calorimeter modules
optimised for electromagnetic and hadronic energy measurements, respectively.

The muon spectrometer (MS) comprises separate trigger and
high-precision tracking chambers measuring the deflection of muons in a magnetic field generated by the superconducting air-core toroidal magnets.
The field integral of the toroids ranges between \num{2.0} and \qty{6.0}{\tesla\metre}
across most of the detector.
Three layers of precision chambers, each consisting of layers of monitored drift tubes, cover the region \(|\eta| < 2.7\),
complemented by cathode-strip chambers in the forward region, where the background is highest.
The muon trigger system covers the range \(|\eta| < 2.4\) with resistive-plate chambers in the barrel, and thin-gap chambers in the endcap regions.

The luminosity is measured mainly by the LUCID--2~\cite{LUCID2} detector that records Cherenkov light produced in the quartz windows of photomultipliers located close to the beampipe.

Events were selected by the first-level trigger system implemented in custom hardware,
followed by selections made by algorithms implemented in software in the high-level trigger~\cite{TRIG-2016-01}.
The first-level trigger accepted events from the \qty{40}{\MHz} bunch crossings at a rate close to \qty{100}{\kHz},
which the high-level trigger further reduced in order to record complete events to disk at about \qty{1.25}{\kHz}.

A software suite~\cite{SOFT-2022-02} %
is used in data simulation, in the reconstruction
and analysis of real and simulated data, in detector operations, and in the trigger and data acquisition
systems of the experiment.


%
\section{Data and simulated event samples}
\label{sec:data_mc}
%

This analysis uses the full LHC Run~2 $pp$ collision dataset collected  during 2015--2018~\cite{DAPR-2018-01},
corresponding to an integrated luminosity of $140~\mathrm{fb}^{-1}$ at a centre-of-mass energy of $\sqrt{s} = 13$~\TeV, with a relative luminosity uncertainty of $0.83\%$~\cite{DAPR-2021-01,LUCID2}.
Events were recorded using a diphoton trigger requiring two electromagnetic clusters with transverse energies $\ET$ above 35 \GeV and 25 \GeV, respectively, both satisfying photon identification requirements based on electromagnetic calorimeter shower-shape variables. The diphoton trigger efficiency relative to the full event selection exceeds 99\% for the 2015–2016 dataset and 98\% for the 2017–2018 dataset~\cite{TRIG-2016-01,TRIG-2018-05,TRIG-2019-03}.

Monte Carlo (MC) simulated events are used to optimise the analysis selections and to characterise the signal and background shapes. Signal event samples are simulated %
using the \MADGRAPH \textsc{lo}~\cite{Alwall:2014hca} (subversion 2.9.9) MC generator for the top partners pair production process and its decay: $pp \rightarrow T\bar{T} \rightarrow S t\, S\bar{t}$. Parton showering and hadronisation are simulated with the A14 tune~\cite{ATL-PHYS-PUB-2014-021} and \NNPDF[2.3lo] set of parton distribution functions (PDF)~\cite{Ball:2012cx} using \PYTHIA[8.307]~\cite{Sjostrand:2007gs}, which is used also for the $S$ decays.
Since this analysis is targeting diphoton final states, one of the spin-0 particles is forced to decay into a pair of photons ($S \rightarrow \gamgam$), while the other one can have any of the possible decays ($S \rightarrow Z\gamma$, $S \rightarrow ZZ$, $S \rightarrow WW$, $S \rightarrow t\bar{t}$, or $S \rightarrow \gamgam$), as all such decays contribute to the overall signal acceptance.
Signal samples are produced for VLQ masses
$m_{T} = 1000$, $1200$, $1400$ and $1800~\GeV$,
and spin-0 particle masses $m_{S} = 200~\GeV$ to $1200~\GeV$, respectively.
The mass grid pairs ($m_{T}$, $m_{S}$) are chosen to provide enough %
coverage to allow reliable interpolation of the signal efficiencies across all channels.
The signals are generated under the narrow-width approximation (NWA) \cite{Banerjee:2022izw,Benbrik:2019zdp}, with $\Gamma/M = 0.1\%$ for both the $T$ and $S$ particles. This ensures that interference effects between the resonant signal and non-resonant contributions, as well as interference with SM background processes leading to the same final state, are negligible.
A small $\Gamma/M$ ratio also implies that single production of $T$ is negligible, since the relevant couplings are very small~\cite{Banerjee:2024zvg}.

The background estimate %
is data-driven, with simulated events used only to motivate the definition of the analysis regions. Background events from continuum diphoton production, neglecting possible interference with a resonant $S \rightarrow \gamgam$ signal, are simulated using \SHERPA~2.2.4~\cite{Bothmann:2019yzt} with the default tuning for the underlying event.
The matrix elements are calculated at next-to-leading-order (NLO) accuracy in the strong coupling constant $\alpha_{\mathrm{S}}$
for up to one additional parton emission and at leading-order (LO) for two and three additional partons,
and are merged with the \SHERPA parton shower according to the ME+PS@NLO prescription~\cite{Hoeche:2012yf}.
The \NNPDF[3.0nnlo]~\cite{Ball:2014uwa} PDF set is used.

Both signal and background events are processed using a fast simulation of the ATLAS detector~\cite{SOFT-2010-01,ATL-PHYS-PUB-2010-013} based on \GEANT~\cite{Agostinelli:2002hh}, where the full simulation of the calorimeter is replaced with a parameterisation of the calorimeter response.
The effect of multiple interactions in the same and neighbouring bunch crossings (\pileup) is modelled by overlaying the simulated hard-scattering event with inelastic proton--proton (\(pp\)) events generated with \PYTHIA[8.186]~\cite{Sjostrand:2007gs} using the \NNPDF[2.3lo] PDF set%
~\cite{Ball:2012cx} and the A3 set of tuned parameters~\cite{ATL-PHYS-PUB-2016-017}.
All simulated events are reconstructed with the same reconstruction algorithms as those used for data.

%
%
%

%


%
\section{Event reconstruction and object selection}
\label{sec:event_object}
%

Events are required to have at least two tracks with \pT larger than 0.5 \GeV associated with one collision vertex~\cite{IDTR-2021-01}.
The hard scatter primary vertex is chosen using a neural network that uses information about photon candidates (see below) and inner detector tracks~\cite{HIGG-2013-08}.

Photon candidates are required to satisfy $\ET^\gamma>38~\GeV$, well above the trigger turn-on threshold, and to be within
\(|\eta| < 1.37\) or \(1.52 < |\eta| < 2.47\).
To reduce the background from jets, photon candidates are required to satisfy \texttt{tight} identification criteria based on the shape of electromagnetic %
showers in the LAr calorimeter and energy leakage into the hadronic calorimeter~\cite{EGAM-2018-01,EGAM-2021-01}.
Events with one or both photon candidates satisfying %
a looser identification are kept for background estimations.
The \texttt{tight} identification is optimised in ranges of %
$\ET^\gamma$ and $|\eta|$.%

To further improve the rejection of jets misidentified as photons, the candidates are required to be isolated using information from both the calorimeter and tracking subsystems.
The calorimeter isolation transverse energy $E^{\text{iso,calo}}_\mathrm{T}$~is required to be smaller than $0.022\ET^\gamma+2.45~\GeV$, where $E^{\text{iso,calo}}_\mathrm{T}$ is defined as the sum of the transverse energies of positive-energy topological clusters~\cite{PERF-2014-07} within a cone of size $\Delta R = 0.2$ around the photon candidate, excluding the photon transverse energy $\ET^\gamma$ and correcting for pile-up and underlying-event contributions~\cite{STDM-2010-08,Cacciari:2008gn,Cacciari:2009dp}.
The track isolation transverse energy $E^{\text{iso,trk}}_\mathrm{T}$~is required to be less than $0.05\ET^\gamma$, where $E^{\text{iso,trk}}_\mathrm{T}$~is defined as the scalar sum of the transverse momenta of tracks with $\pT > 1$~\GeV\ in a $\Delta R=0.2$ cone around the photon candidate.
The tracks are required to satisfy some loose track-quality criteria, not to be associated with a photon conversion, and compatible with the diphoton production vertex.

Jets are reconstructed using the \antikt\ algorithm~\cite{Cacciari:2008gp} as implemented in FastJet~\cite{Fastjet} with a jet radius parameter $R = 0.4$.
The inputs to this algorithm are particle flow (PFlow)  objects~\cite{PERF-2015-09}, which combine measurements from the ATLAS inner detector and calorimeters
to improve the jet energy resolution and increase the jet reconstruction efficiency, especially at low jet \pt.
Jets are required to have $\pt>28~\GeV$ and $|\eta| < 4.4$, and must satisfy the standard jet-quality and cleaning criteria.
Calibrations are applied to the jet energy scale (JES) and jet energy resolution (JER), which include components derived from both simulation and in situ measurements~\cite{JETM-2018-05}.
Jets originating from $b$-quarks, $b$-jets, are identified within $|\eta|<2.5$ using the DL1r algorithm~\cite{FTAG-2019-07}  %
with a fixed 70\% efficiency working point.

Electrons are required to have $\pt>10~\GeV$ and %
\(|\eta| < 1.37\) or \(1.52 < |\eta| < 2.47\), which excludes the crack region, and to satisfy the \texttt{Medium} identification working point and \texttt{Loose} isolation criteria~\cite{EGAM-2018-01}, as well as standard track-to-vertex association requirements.
Muons are reconstructed by combining inner detector and muon spectrometer tracks, and must satisfy $\pt>10~\GeV$, $|\eta|<2.7$, the \texttt{Medium} identification, and the \texttt{Loose} isolation working point.
Standard impact-parameter significance and longitudinal distance cuts are applied~\cite{MUON-2018-03}. %

The missing transverse momentum is defined as the negative vector transverse momentum sum of the reconstructed and calibrated physics objects, plus an additional soft term~\cite{JETM-2020-03}.
The soft term is built from tracks that are not associated with any reconstructed electron, muon or jet, but which are associated with the primary vertex.
The magnitude of the missing transverse momentum is denoted by~\MET.

An overlap removal procedure is applied to prevent double counting of physics objects, following the same sequence as in a previous ATLAS high-mass diphoton resonance search~\cite{HIGG-2018-27}. Objects are removed in the following order using the angular separation $\Delta R=\sqrt{(\Delta\eta)^2+(\Delta\phi)^2}$: electrons close to photons ($\Delta R<0.4$), muons close to photons ($\Delta R<0.4$), jets close to photons ($\Delta R<0.4$), jets close to electrons ($\Delta R<0.2$), electrons close to jets ($\Delta R<0.4$), and muons close to jets ($\Delta R<0.4$). %

The main discriminant observable of the analysis is defined as the scalar sum of the transverse momenta of jets, leptons and photons, denoted by $\Ht$, plus~\MET. This choice is motivated by the topology expected for the pair production of heavy resonances: given their large mass, their decay products are typically emitted with large momenta projections in the transverse plane and ~\HtMET provides an estimate of the mass scale of the underlying resonances.
Furthermore, large values of the total transverse energy %
are strongly suppressed in any Standard Model process, making~\HtMET a powerful discriminant. Although its distribution for the signal is broad, it is well separated from the SM background, which populates only the lower end of the spectrum.

The analysis selection requires at least two photons and at least one $b$-jet satisfying the criteria described previously, together with a large value of \HtMET. Only events with an invariant mass of the diphoton system, $\mgg$ larger than 150~\GeV are considered. This threshold is chosen to avoid overlap with the Higgs boson mass region.
If there are more than two photons fulfilling the criteria in the event, the two with highest \pt are selected.

\begin{figure}[htbp]
\centering
\includegraphics[width=0.65\textwidth]{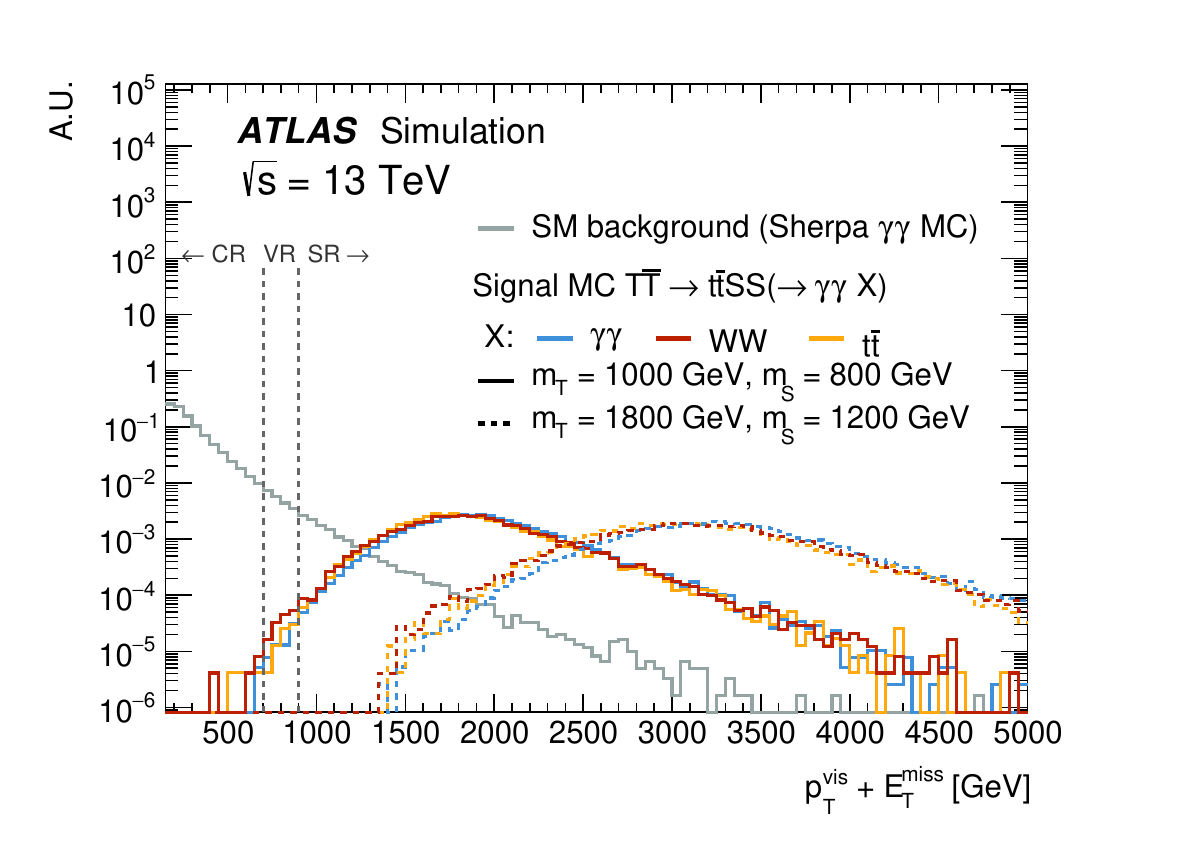}
\caption{
Distributions of \HtMET for representative signal samples in the $4\gamma, t\bar{t}$ and $WW$  final states and the SM background simulated events from continuum \gamgam production. Control, validation and signal regions (CR,VR and SR respectively) are delimited by dashed vertical lines.
\label{fig:HT_Emiss}}
\end{figure}

Figure~\ref{fig:HT_Emiss} shows the \HtMET distribution for the simulated signal samples and the continuum $\gamgam$ background, demonstrating the strong discriminating power of this variable.

Various orthogonal slices of $\HtMET$ are used to define control regions (CRs) and signal regions (SRs), as specified in Table~\ref{table:regions}. The division of control regions is made to provide various background templates to cover for any dependence on $\HtMET$, while the SRs are designed to be sensitive to a range of VLQ masses.
The validation region (VR) is defined at intermediate values of $\HtMET$, between the control and signal regions with negligible signal contamination. It serves to test the modelling of backgrounds in a kinematic regime closer to the SR, in particular the choice of the background modelling functional form before unblinding data, providing a cross-check that the background modelling from the control regions is reliable before examining the SRs.

\begin{table}[h!tbp]
\renewcommand{\arraystretch}{1.25}
\caption{Definition of regions based on $\HtMET$ cuts and observed event yields.}
\label{table:regions}
\begin{center}
\begin{tabular}{ccc}
\hline
\rule{0pt}{2.3ex}Region & $\HtMET$ range & Data event yield  \\
\hline
\rule{0pt}{2.3ex}Control Region 1 (CR1) & $300~\GeV < \HtMET \leq 400~\GeV$ & 2892 \\
\rule{0pt}{2.3ex}Control Region 2 (CR2) & $400~\GeV < \HtMET \leq 500~\GeV$ & 1847 \\
\rule{0pt}{2.3ex}Control Region 3 (CR3) & $500~\GeV < \HtMET \leq 600~\GeV$ & 1168 \\
\rule{0pt}{2.3ex}Control Region 4 (CR4) & $600~\GeV < \HtMET \leq 700~\GeV$ & 732 \\
\hline
\rule{0pt}{2.3ex}Validation Region (VR) & $700~\GeV < \HtMET \leq 900~\GeV$ & 785 \\
\hline
\rule{0pt}{2.3ex}Signal Region 1 (SR1) & $900~\GeV < \HtMET \leq 1100~\GeV$ & 312 \\
\rule{0pt}{2.3ex}Signal Region 2 (SR2) & $1100~\GeV < \HtMET \leq 1400~\GeV$ & 196 \\
\rule{0pt}{2.3ex}Signal Region 3 (SR3) & $1400~\GeV < \HtMET \leq 2000~\GeV$ & 104 \\
\rule{0pt}{2.3ex}Signal Region 4 (SR4) & $\HtMET > 2000~\GeV$ & 26 \\
\hline
\end{tabular}
\end{center}
\end{table}

To provide a more model-independent result, following the detector-level selection, the fit to %
the signal production cross-section is performed in a fiducial volume defined from the simulated samples by requiring at least two photons at particle level with $\ET>38$~\GeV, $|\eta|<2.37$ and $m_{\gamma\gamma}$ to be within $0.5\%$ of the resonance mass $m_{S}$, which corresponds to five times the simulated signal width.


%
\section{Signal and background modelling}
\label{sec:signal_model}
%

%

%

%

The resonant signals and the different backgrounds are separated using the diphoton invariant mass, $m_{\gamma\gamma}$, as the discriminating observable through a fit using analytical functions for both components.
The shape of the diphoton signal invariant mass distribution for each signal hypothesis is modelled with a double-sided Crystal Ball (DSCB) function for all SRs, which is composed of a Gaussian distribution core with power-law tails~\cite{HIGG-2014-04,Oreglia:1980cs}, whose parameter values evolve linearly relative to the mass~\mS.
The parameters of the DSCB function are extracted from fits to simulated signal samples for each channel and $m_{T}$ masses.
The width of the Gaussian function core is entirely determined by the resolution of the detector and ranges from 1.8 to 10.2~\GeV for $m_{S}$ masses between 200 and 1200~\GeV and it is independent of the final state in which it is measured.

Acceptable %
agreement between the signal parameterisation and the simulated signal points is found for all final states, with differences below 2\% of the fitted signal yield for channels with two photons in the final state and up to 10\% in the $\gamgam\gamgam$ and $\gamgam Z\gamma$ final states.
The analysis is statistically limited, and the impact of this systematic uncertainty on the final result is negligible compared with the total uncertainty, see Table~\ref{tab:systs} in Section~\ref{sec:stats}.

%
%
%
%
%
%
%
%
%
%
%
%
%
%
%


%
%
\label{sec:bkg_model}
%


The analysis makes use of a data-driven background estimate in which
the continuum background shape in each SR is parameterised by an analytic function, whose parameters are left free in the fit to data in all regions. The diphoton invariant mass distribution of each data CR provides a background template orthogonal to the SRs by construction, as depicted in Table~\ref{table:regions}. This provides an unbiased estimate of the continuum background composed of non-resonant quantum chromodynamic (QCD) diphoton production and events in which one or both photons are misidentified electrons or jets, predominantly the latter. The flexibility of the chosen functional form is then tested in the VR, since it provides an invariant diphoton mass distribution in a region closer to the SRs. The uncertainty arising from the choice of background model is based on signal-plus-background fits following the methodology described in Ref.~\cite{ATL-PHYS-PUB-2020-028}, and for which all background templates from CRs and VR are considered as variations of the background shape.

In all regions, the diphoton invariant mass distribution for the SM background is described using an exponential function of the form
\begin{equation}
f(m_{\gamma\gamma}) = \mathcal{N}\, e^{-m_{\gamma\gamma}/\tau},
\end{equation}
that %
contains a single free shape parameter, $\tau$.
The chosen function provides good fits in all CRs and in the VR,
with $p$-values ranging from $2\%$ to $50\%$ .
Functional forms with higher orders in the exponent were considered but did not lead to an improvement in the background description.

\begin{figure}[htbp]
\centering
\subfloat[]{
\includegraphics[width=0.48\linewidth]{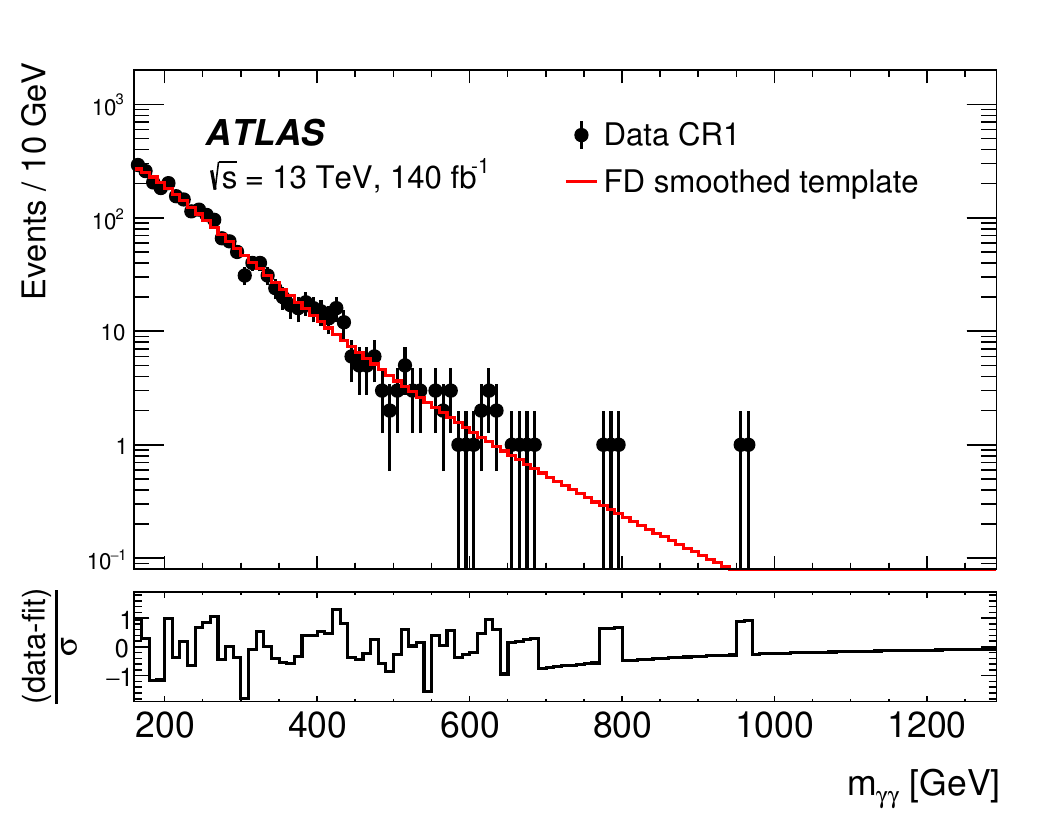}\label{fig:fdPlot}}
\subfloat[]{
\includegraphics[width=0.5\linewidth]{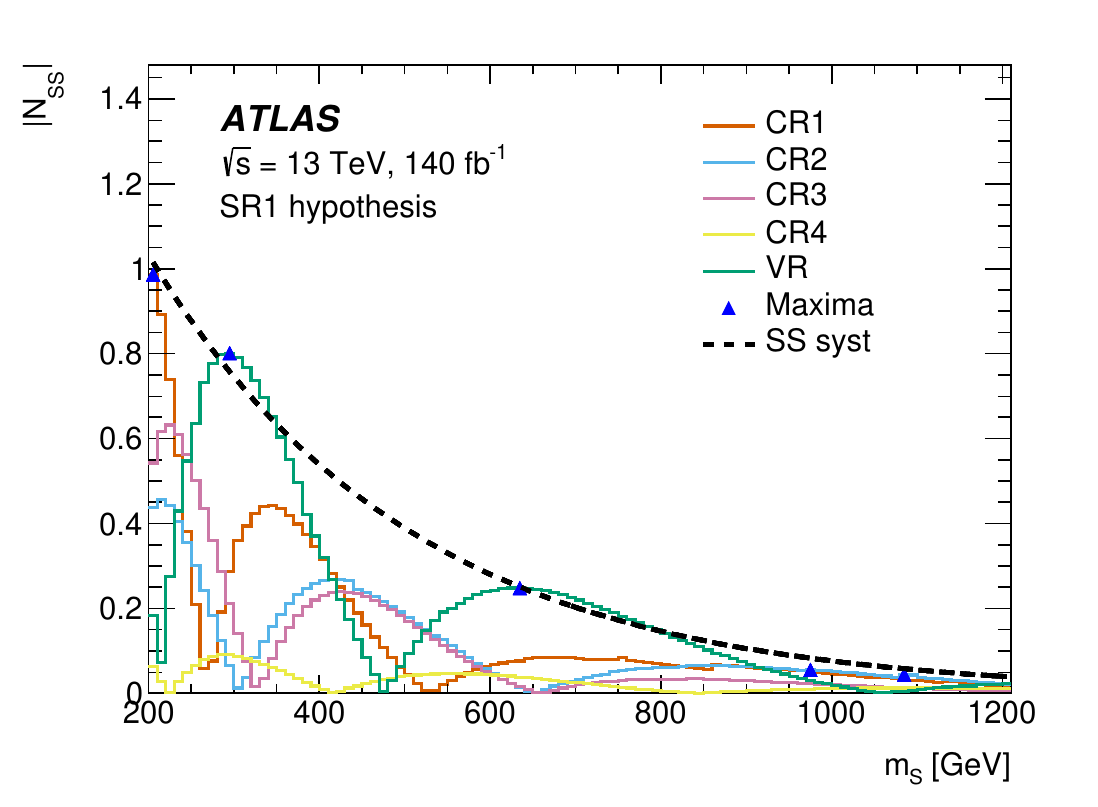}\label{fig:ssAbsPlot}} \\
\caption{(a) An example of the functional decomposition (FD) fit to the $m_{\gamma\gamma}$ spectrum of the data in CR1 where the residual panel shows the residuals of data with respect to the smoothed template. (b) Absolute number of spurious signal events $|N_{SS}|$ as a function of $m_{S}$ in SR1 evaluated for the smoothed templates obtained from each CR and VR\@. The %
triangles show the maxima used for obtaining the spurious signal systematic uncertainty (%
dashed line). }
\end{figure}

The bias arising from the choice of the background model in each SR is evaluated from signal-plus-background fits on background-only templates: any fitted signal yield $N_{SS}$ is designated  ``spurious signal'' (SS) and is considered as a systematic uncertainty in the modelling of the background shape.
The estimate of this systematic uncertainty requires the shape of the background template to be as close as possible to the data but with an event count large enough so that the statistical uncertainties of the template are negligible.
To meet the %
requirements, a smoothing of the background templates is performed using the functional decomposition (FD) method~\cite{Edgar:2018irz} following the prescription described in Ref.~\cite{HIGG-2018-27}.
An example of the obtained smoothed template from the data in CR1 is shown in Figure~\ref{fig:fdPlot}.

The normalisation of each smoothed template, obtained from each CR and including the VR, is scaled to the observed yield in each SR and then fit with the background model together with the signal model. This procedure is performed for all diphoton resonance mass hypotheses $m_S$ in the search range. The measured spurious signal $N_{SS}$ remains below $30\%$ relative to the statistical uncertainty of the signal yield across the entire mass range for SR1 and decreases for subsequent SRs. Figure~\ref{fig:ssAbsPlot} shows the spurious signal systematic uncertainty for all the CRs and VR assuming the event yield in SR1. The final background modelling uncertainty for each is computed as the envelope of the maximum fitted yields over all background templates and is parameterised with a decreasing exponential function.

%
\section{Statistical analysis}

\label{sec:stats}

The statistical interpretation of this search is based on an unbinned simultaneous likelihood fit to the four SRs. Each SR is modelled using a profile-likelihood approach, where signal and background yields are extracted from fits to the diphoton invariant mass distribution, incorporating systematic uncertainties via nuisance parameters. Since these regions are orthogonal by construction, the likelihood is built by multiplying their individual likelihoods, with systematic uncertainties treated as fully correlated through a single Gaussian constraint, except for the background modelling uncertainty that %
is implemented as an additional signal-like component to account for potential spurious signal contributions.

The parameter of interest to be extracted from the likelihood fit is the fiducial production cross-section $\sigma_{\text{fid}}(pp\rightarrow T\bar{T})$  times the branching ratios related to the channel under consideration:
\begin{equation*}
\sigma_{\text{fid}}(pp\rightarrow T\bar{T}) \cdot \mathcal{B}(S\rightarrow \gamgam)\cdot\mathcal{B}(S\rightarrow \text{X}) = \frac{N_{S}}{\alpha C^{\gamgam \text{X}}_{S}\mathcal{L}}\text{, with } C^{\gamgam \text{X}}_{S} = \frac{N^{\text{reco},\gamgam \text{X}}_{\text{MC}}}{N^{\text{fid},\gamgam \text{X}}_{\text{MC}}},
\end{equation*}
where $N_{S}$ is the number of signal events fitted in data, $\mathcal{L}$ is the integrated luminosity, $N^{\text{reco},\gamgam \text{X}}_{\text{MC}}$ is the number of reconstructed and selected signal events in the simulation in a given channel $\gamgam \text{X}$ (with $X \in \{\gamgam, \gamma Z, WW,ZZ,t\overline{t}\}$) and analogously, $N^{\text{fid},\gamgam \text{X}}_{\text{MC}}$ is the number of simulated signal events present within the fiducial volume.
The constant $\alpha$ is a factor that accounts for the indistinguishability of the decays of both $S$ particles in the $4\gamma$ channel and therefore it is equal to unity for that channel and equal to two elsewhere.
The decay branching ratio for the top partner $T$ is taken as $\mathcal{B}(T\rightarrow St) = 100\%$.

To provide continuous coverage of the $(m_T,m_S)$ parameter space, maps of the correction factors $C^{\gamgam \text{X}}_{S}$ are obtained using radial basis function (RBF) interpolation with a multi-quadric kernel~\cite{Rasmussen2006}, together with a logit transformation to preserve positivity. These maps, evaluated separately for each SR and decay channel, are used to estimate the expected signal yields and interpolate the upper limits between the simulated mass points. The correction factors reach values of up to approximately 50\%, depending on the channel, with the highest efficiencies observed in SR4, while decreasing to values four orders of magnitude smaller in SR1. A systematic uncertainty is assigned to the interpolated efficiencies by assessing the stability of the interpolated surfaces under variations of the RBF hyperparameters.

The theoretical uncertainties affecting this measurement arise from variations of the renormalisation and factorisation scales affecting the signal efficiencies evaluated in simulated samples.
The experimental uncertainties directly impacting the signal yield include those involved in the luminosity determination, the modelling of pile-up interactions in simulation, the trigger efficiency, and photon identification and isolation selection efficiencies.
Uncertainties in the signal shape parameterisation from the modelling and the determination of the photon
energy resolution and scale are also accounted for. The overall impact of jet-related, lepton and \met{} systematic uncertainties are negligible on the signal efficiency and therefore are not accounted for in the fit.
All sources of systematic uncertainties are summarised in Table~\ref{tab:systs}.

\begin{table}[!ht]
\centering
\caption{Summary of the sources of systematic uncertainty considered. %
Their impact on the fiducial production cross-section of a hypothetical resonant signal is shown, except for the background modelling uncertainty, which is expressed both as number of events and relative to the expected statistical uncertainty $\delta S$ of a fitted signal.
Unless written otherwise, numbers are mass and channel independent. The ``Remarks'' column indicates specific information about the systematic uncertainty, including whether or not the uncertainty varies as a function of resonance mass or SR.
}
\begin{tabular}{lcc}
\hline
Source                   & Uncertainty  & Remarks\\
\hline
\multicolumn{3}{c}{On $\sigma_{\text{fid}}(pp\rightarrow T\bar{T})\cdot \mathcal{B}(S\rightarrow \gamgam)\cdot\mathcal{B}(S\rightarrow X)$}                                           \\
\hline
Efficiency interpolation & 1--15$\%$ & channel and SR-dependent\\
Signal shape modelling $N_\gamma>2$ & $10\%$& for $\gamgam\gamgam$ and \gamgam$Z\gamma$\ channels \\
Signal shape modelling $N_\gamma=2$& ${<}1\%$& for all channels with 2 photons \\
Photon energy resolution & $^{+7.6\%}_{-2.3\%}$ -- $^{+4.0\%}_{-1.3\%}$ &  $m_{S}$-dependent \\
Scale and PDFs uncertainties & $\pm 3.0 \%$ & \\
Isolation efficiency     & $\pm 3.0 \%$ &\\
Photon identification    & $\pm 2.0 \%$ & \\
Trigger                  & $\pm 1.2\%$ &\\
Luminosity  & $\pm 0.83 \%$ &\\ %
Pile-up modelling & $\pm  0.1\%$ & \\

\hline
\multicolumn{3}{c}{\em Background modelling}                                           \\
\hline
Spurious signal & 1.03 -- <0.01 events (30\% -- <1\%) & $m_{S}$ and SR-dependent  \\
\hline
\end{tabular}
\label{tab:systs}
\end{table}


%
\section{Results}
\label{sec:results}
%

\begin{figure}[htbp]
\centering
\subfloat[]{

\includegraphics[width=0.49\linewidth]{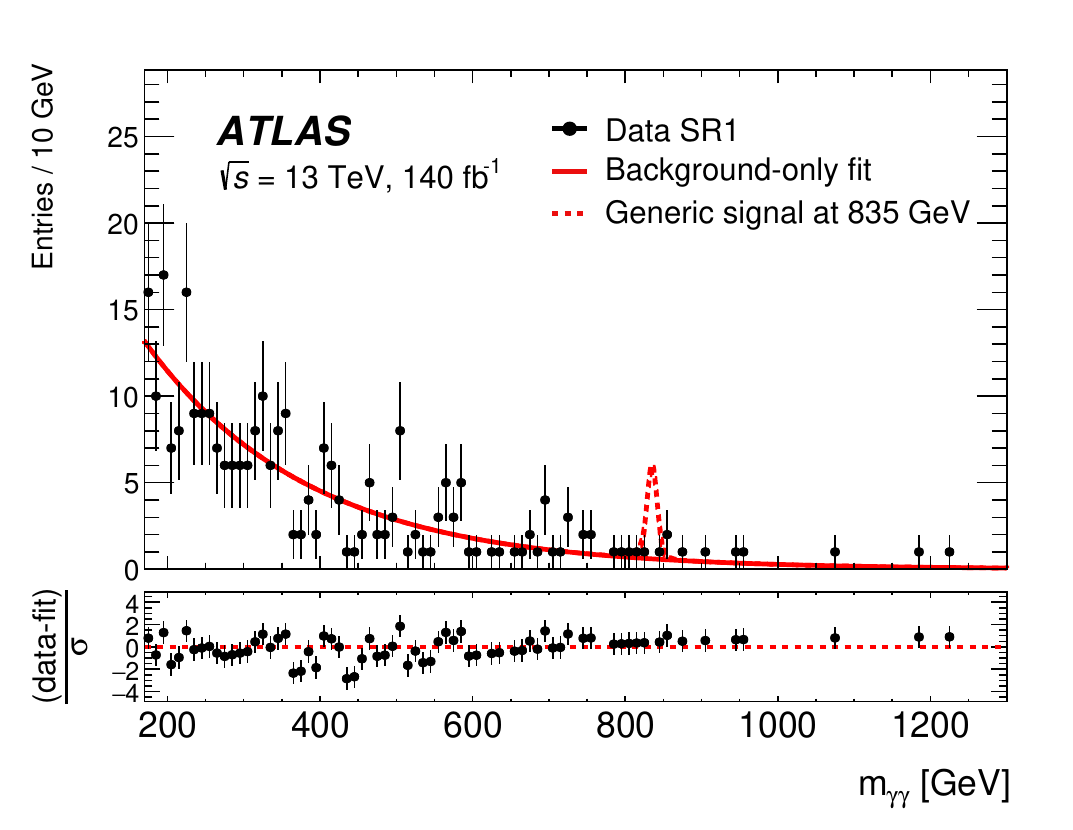}}
\subfloat[]{

\includegraphics[width=0.49\linewidth]{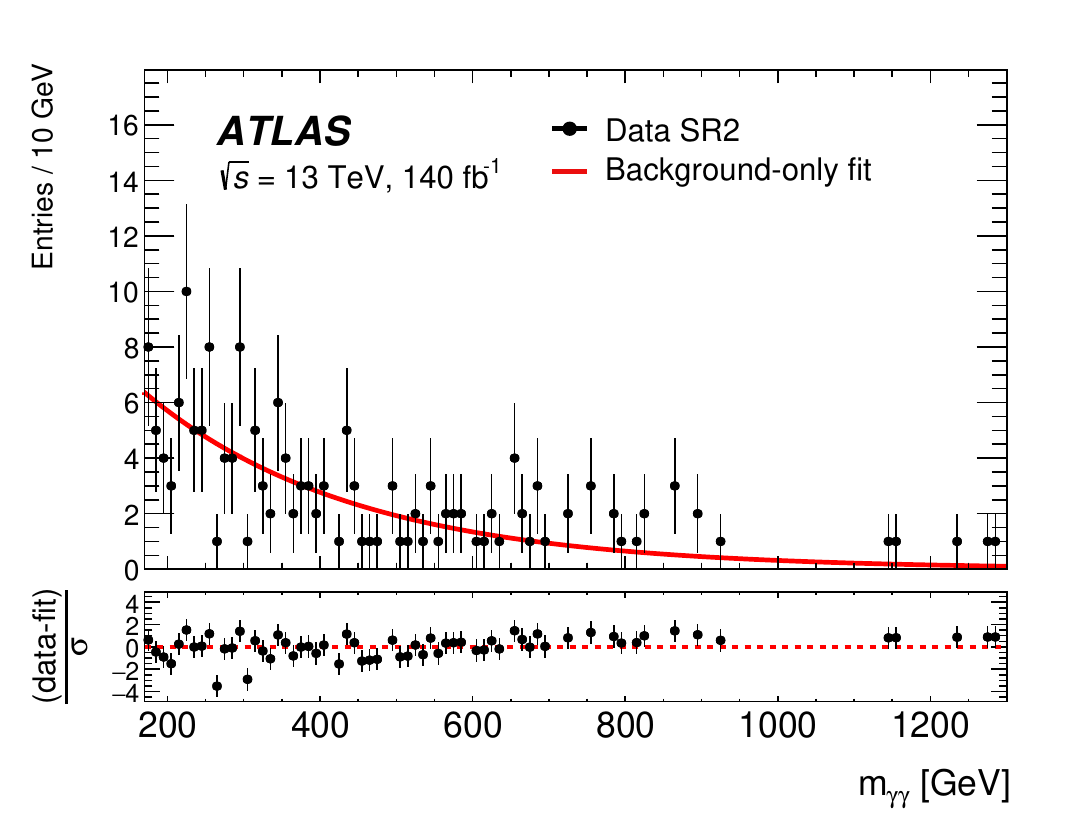}}\\
\subfloat[]{

\includegraphics[width=0.49\linewidth]{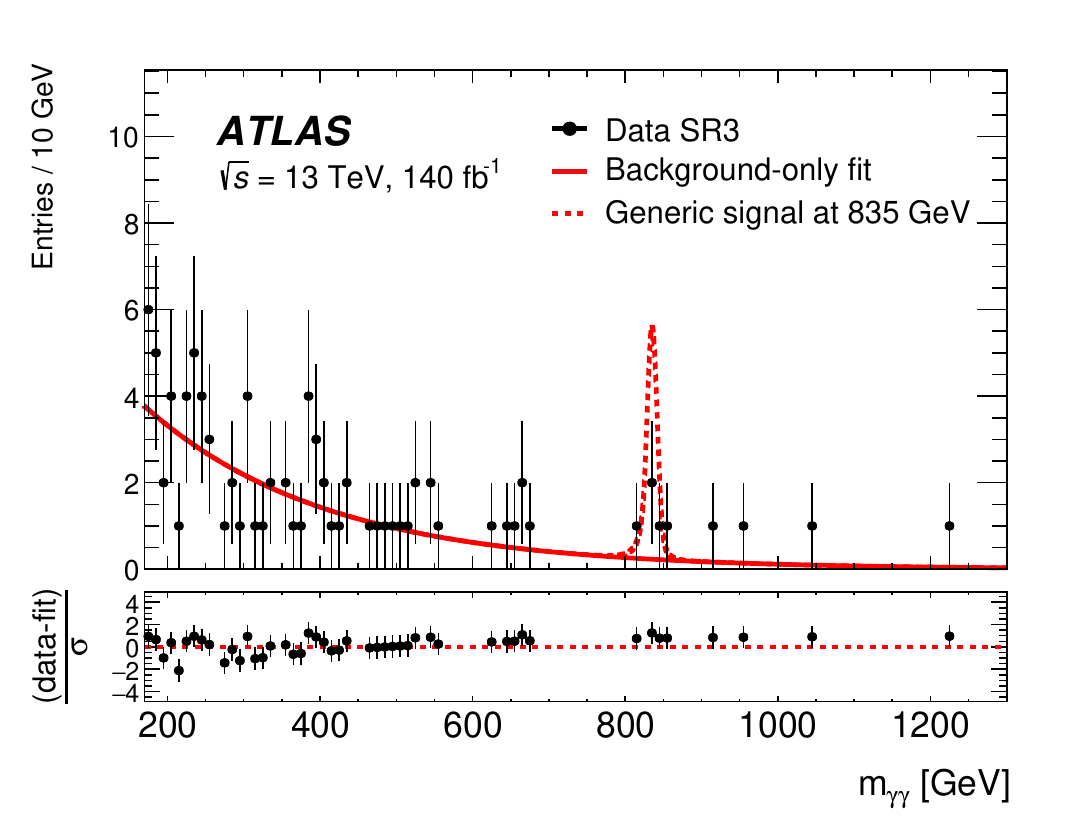}}
\subfloat[]{

\includegraphics[width=0.49\linewidth]{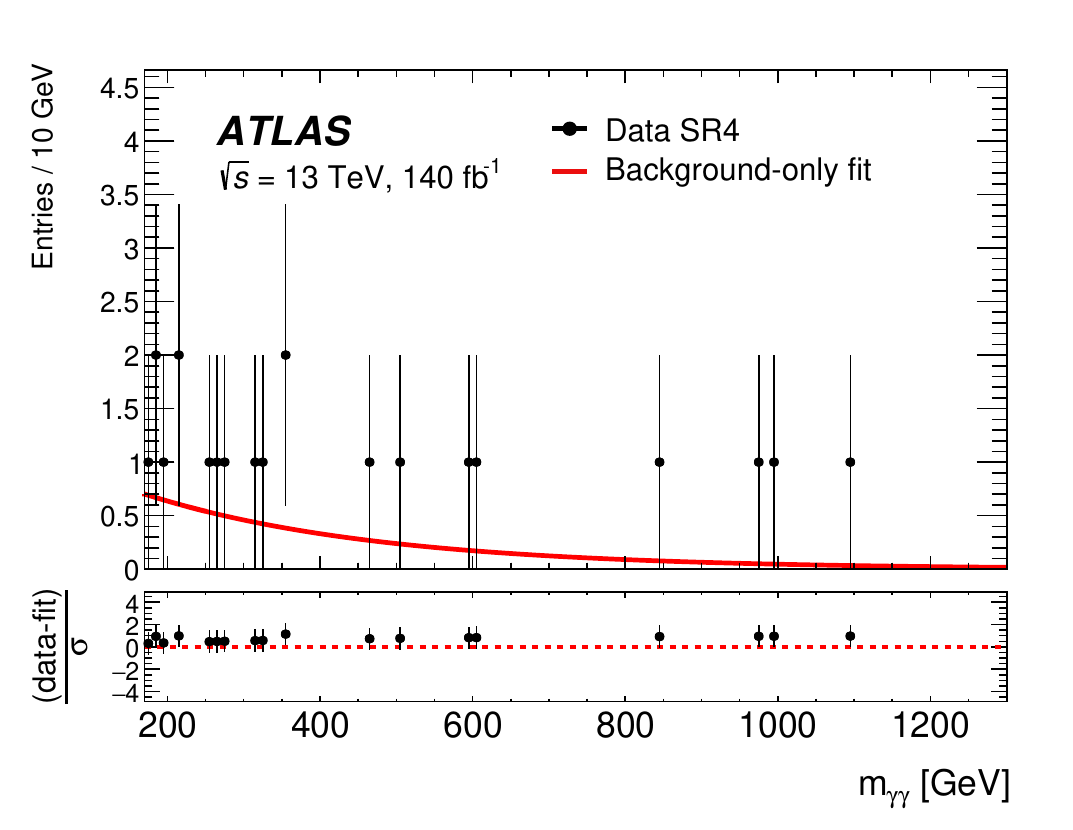}}\\
\caption{Distribution of the diphoton invariant mass for all events satisfying the analysis selections for each of (a) SR1, (b) SR2, (c) SR3, and (d) SR4 in the full Run 2 dataset. The background-only fit to each of the SRs is shown
as a solid line while dashed lines show a generic signal shape.
\label{fig:datafitSR}}
\end{figure}

The diphoton invariant mass distribution of the events that satisfy the selection in the four SRs is shown in Figure~\ref{fig:datafitSR}, along with the background only fit using the functional form described in Section~\ref{sec:bkg_model}.

The compatibility of the observed data and the background-only hypothesis for a given signal hypothesis is tested by estimating a local $p$-value based on a profile-likelihood-ratio test statistic, detailed in Ref.~\cite{Cowan:2010js}.
The data are overall consistent with the background estimate, with deviations within $2\sigma$.%

In the absence of a signal, the expected and observed 95\% confidence level (CL) exclusion limits on the cross-section times branching ratio are evaluated using the modified frequentist approach CL$_\text{s}$~\cite{Junk:1999kv,Read:2002hq} with the asymptotic approximation~\cite{Cowan:2010js} to the test-statistic distribution.

\begin{figure}[h!]
\centering
\subfloat[]{\includegraphics[width=0.47\columnwidth]{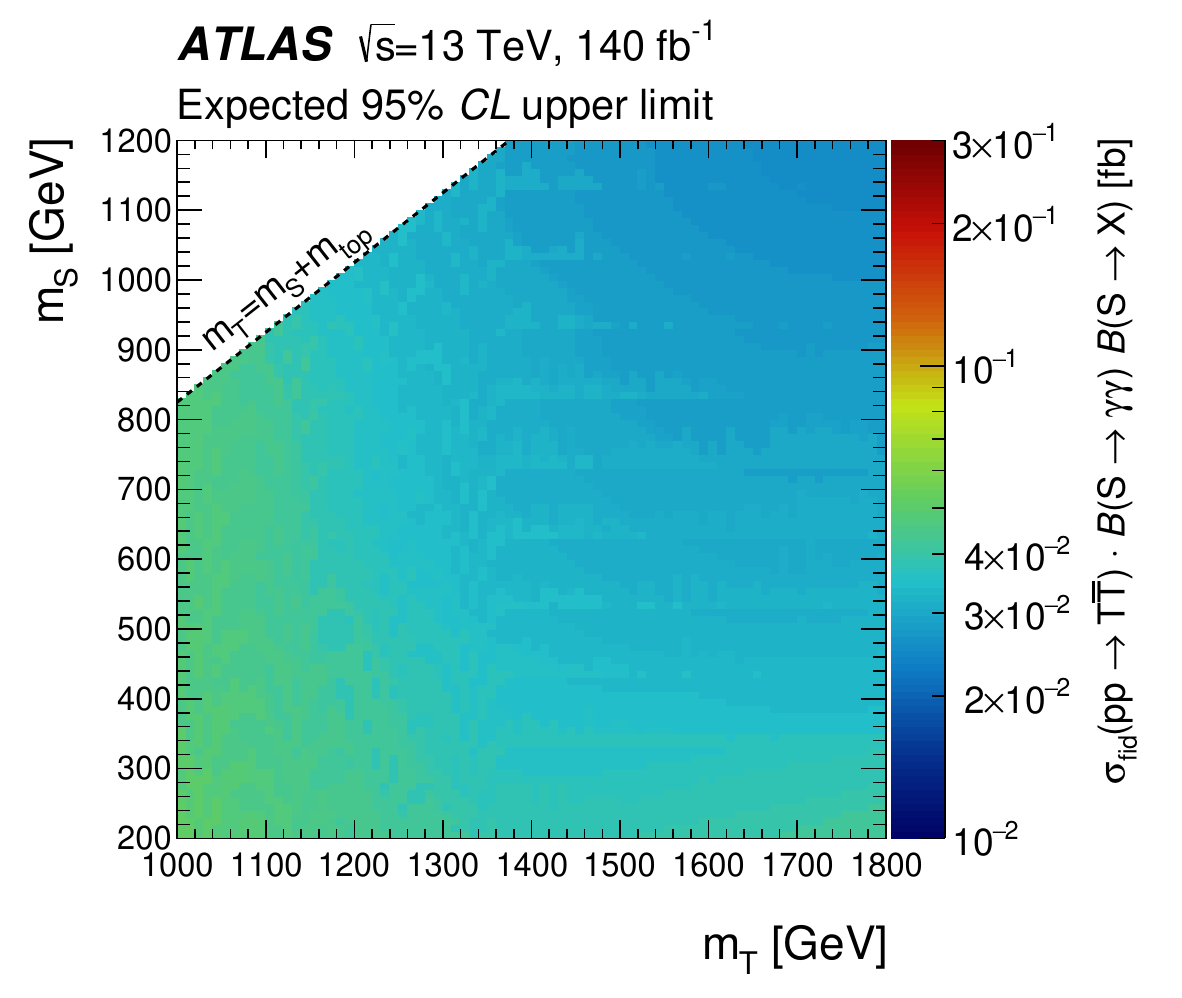}}
\subfloat[]{\includegraphics[width=0.47\columnwidth]{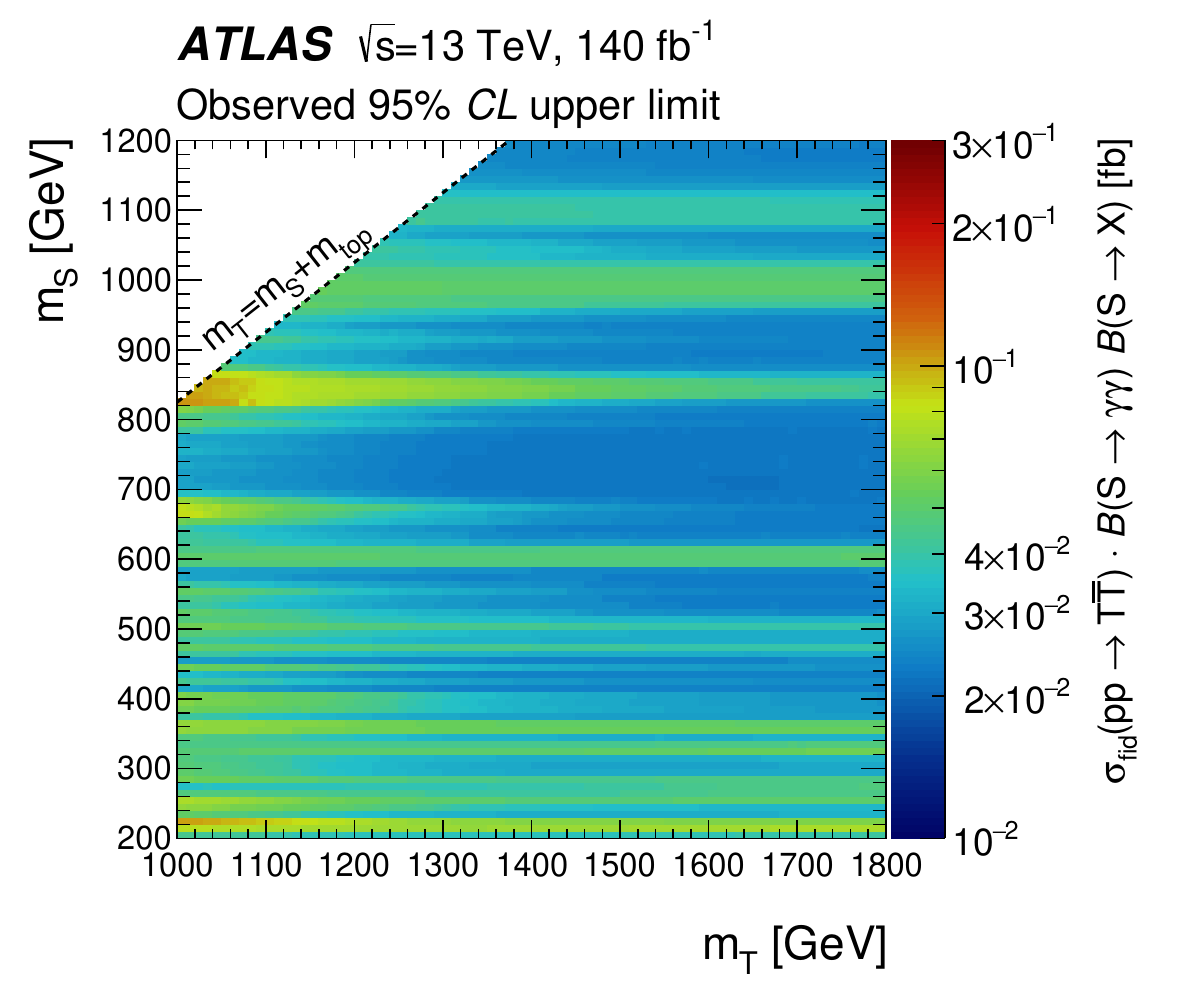}}\\
\caption{Expected (a) and observed (b) 95\% CL upper limits on the fiducial production cross-section times branching ratio as a function of the vector-like quark $T$ mass and the spin-0 particle mass $(m_T, m_S)$. The best obtained limit for any channel is shown for each mass point.
\label{fig:bestlimit_plots}}
\end{figure}

Two-dimensional upper limits on the fiducial cross-section times branching ratio in the $(m_T, m_S)$ plane are computed for all channels ($\gamgam\gamgam$, $\gamgam ZZ$, $\gamgam Z\gamma$, $\gamgam \ttbar$, and $\gamgam WW$) and the best cross-section limit for each mass point and channel is shown in Figure~\ref{fig:bestlimit_plots}.
Given the nature of the analysis, the upper limits on the fiducial cross-section are very similar for all channels with three or two photons  (i.e.~for the $\gamgam X$ channels, where $X \in \{\gamma Z, WW, ZZ, \ttbar \}$). \footnote{A factor of 2 larger upper limits are observed for the $\gamma\gamma\gamma\gamma$ final state caused by the indistinguishability of the final state.}
For the same reason, these limits also hold for any $\gamgam +X$ decay channel, not limited to the ones considered here, where $X$ contains one photon or fewer.
As stated before, given the model independent approach within the fiducial volume, the obtained upper limits can be reinterpreted in a wide variety of phenomenological models. %

The obtained results are interpreted in the framework of the model proposed in Ref.~\cite{Banerjee:2022izw}.
This model contains various vector-like quarks of different charges, including two nearly-degenerate top partners, $T_{2/3}$ and $X_{2/3}$.
These VLQs can decay into three neutral pseudoscalars $\eta$, $\chi^0_1$ and $\chi^0_5$, all represented by $S$, and a top quark.
Other decays into SM particles and charged scalars are allowed but not considered in the following. The pseudoscalars subsequently decay into different final states with specific and mass-dependent branching ratios, as reported in Ref.~\cite{Banerjee:2022izw}.
The reinterpretation is performed by summing the signal coming from each pair-produced VLQ and leading to final states with $t\bar t$ and $4\gamma$, $3\gamma+X$ and $2\gamma+X$, where $X$ are any allowed SM particles.
This procedure is simplified with respect to the calculation of Ref.~\cite{Banerjee:2022izw}: the two VLQs are assumed degenerate in mass, and also the four spin-0 particles are assumed to be degenerate among themselves.

An effective cross-section incorporating all the contributions, including the $T\bar{T}$ cross-section, to a given channel is computed and compared with the upper limits in the ($m_T$--$m_S$) plane, leading to a combined bound on the theoretical model.
\begin{figure}[htbp]
\centering
\includegraphics[width=0.65\textwidth]{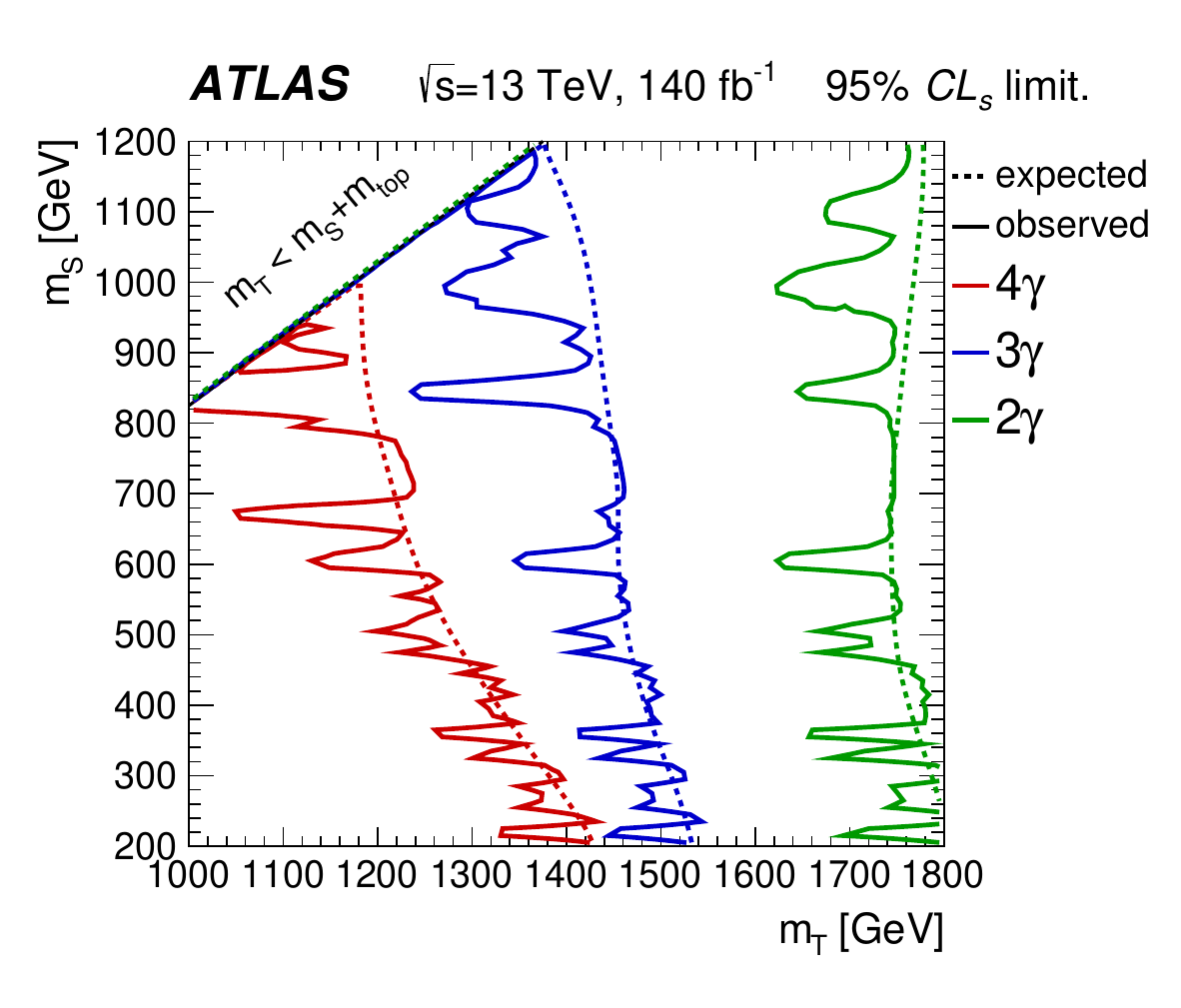}
\caption{Observed (solid) and expected (dashed) 95\% CL exclusion contours in the ($m_{T}$, $m_{S}$) plane for the $4\gamma$, $3\gamma$, and $2\gamma$ cases on the product of the pair production cross-section times branching ratios of the two vector-like top partners and the four spin-0 particles of the model in Ref.~\cite{Banerjee:2022izw}, assuming that the VLQs are degenerate in mass and the spin-0 particles are also degenerate among themselves.
The area to left of each curve is is excluded. The diagonal boundary in the upper-left corner marks the kinematically forbidden region $m_{T} < m_{S} + m_{\text{top}}$, which is excluded from consideration independent of the analysis sensitivity.
\label{fig:interpration}}
\end{figure}
The $T\bar{T}$ cross-section, which is a QCD process, is computed with \HATHOR~\cite{Aliev:2010zk} and NNPDF 4.0 \textsc{nnlo}~\cite{NNPDF:2021njg}.

Observed and expected upper exclusion limits on the product of the pair-production cross-section and branching ratios, using the model from Ref.~\cite{Banerjee:2022izw}, are shown in Figure~\ref{fig:interpration}.  The $4\gamma$, $3\gamma$, and $2\gamma$ curves correspond to the scenarios in which the two spin-0 particles decay into $\gam\gam + \gam\gam $, $\gam\gam + Z \gam$  and $\gam\gam$ plus a non-photonic final state, respectively. Comparing the three curves illustrates the sensitivity of the search under different assumptions on the scalar decay branching ratios. The strongest exclusion is seen for the $2\gamma$ scenario, where VLQ masses less than $m_{T}=1620~\GeV$ are excluded over the full $m_{S}$ range considered.


%
\FloatBarrier
\section{Conclusion}
\label{sec:conclusion}
%

This paper presents the first ATLAS search for non-standard decays of VLQs, specifically targeting pair-production of vector-like $T$ quarks, where one $T$ decays like $T \rightarrow St \rightarrow \gamma\gamma t$ and the other decays like $T \rightarrow St$, with $S \rightarrow \gamma\gamma$, $S\rightarrow Z\gamma$, $S\rightarrow \ttbar$, $S\rightarrow WW$ or $S\rightarrow ZZ$.
These results address a previously unexplored search topology in our understanding of VLQs, as no previous study has searched for VLQ pairs decaying via exotic scalars or pseudoscalars. The search uses the full Run 2 dataset and explores the phase space in $(\mT, \mS)$ that spans $[1000,1800] \times [200,1200]$~\GeV, excluding kinematically suppressed regions.

No significant excess above the expected background is observed for any of the signal hypotheses considered.
The results are also presented in terms of upper limits on the fiducial production cross-section of the VLQ and are further interpreted to constrain the model of Ref.~\cite{Banerjee:2022izw}.
In this model, $T$ masses less than $m_{T} = 1620~\GeV$ for all $m_{S}$ range considered  are excluded for the $2\gamma$ case.


%
%
%
%
%
\section*{Acknowledgements}
%

%

%
%

%
%

We thank CERN for the very successful operation of the LHC and its injectors, as well as the support staff at
CERN and at our institutions worldwide without whom ATLAS could not be operated efficiently.

The crucial computing support from all WLCG partners is acknowledged gratefully, in particular from CERN, the ATLAS Tier-1 facilities at TRIUMF/SFU (Canada), NDGF (Denmark, Norway, Sweden), CC-IN2P3 (France), KIT/GridKA (Germany), INFN-CNAF (Italy), NL-T1 (Netherlands), PIC (Spain), RAL (UK) and BNL (USA), the Tier-2 facilities worldwide and large non-WLCG resource providers. Major contributors of computing resources are listed in Ref.~\cite{ATL-SOFT-PUB-2026-001}.

We gratefully acknowledge the support of ANPCyT, Argentina; YerPhI, Armenia; ARC, Australia; BMWFW and FWF, Austria; ANAS, Azerbaijan; CNPq and FAPESP, Brazil; NSERC, NRC and CFI, Canada; CERN; ANID, Chile; CAS, MOST, NSFC and FRFCU, China; Minciencias, Colombia; MEYS CR, Czech Republic; DNRF and DNSRC, Denmark; IN2P3-CNRS and CEA-DRF/IRFU, France; SRNSFG, Georgia; BMFTR, HGF and MPG, Germany; GSRI, Greece; RGC and Hong Kong SAR, China; ICHEP and Academy of Sciences and Humanities, Israel; INFN, Italy; MEXT and JSPS, Japan; CNRST, Morocco; NWO, Netherlands; RCN, Norway; MNiSW, Poland; FCT, Portugal; MNE/IFA, Romania; MSTDI, Serbia; MSSR, Slovakia; ARIS and MVZI, Slovenia; DSI/NRF, South Africa; MICIU/AEI, Spain; SRC and Wallenberg Foundation, Sweden; SERI, SNSF and Cantons of Bern and Geneva, Switzerland; NSTC, Taipei; TENMAK, T\"urkiye; STFC/UKRI, United Kingdom; DOE and NSF, United States of America.

Individual groups and members have received support from BCKDF, CANARIE, CRC and DRAC, Canada; CERN-CZ and PRIMUS, Czech Republic; COST, ERC, ERDF, Horizon 2020 and Marie Sk{\l}odowska-Curie Actions, European Union; Investissements d'Avenir Labex, Investissements d'Avenir Idex and ANR, France; DFG and AvH Foundation, Germany; Herakleitos, Thales and Aristeia programmes co-financed by EU-ESF and the Greek NSRF, Greece; BSF-NSF and MINERVA, Israel; NCN and NAWA, Poland; La Caixa Banking Foundation, CERCA and AGAUR programs from Generalitat de Catalunya and PROMETEO and GenT Programmes Generalitat Valenciana, Spain; G\"{o}ran Gustafssons Stiftelse, Sweden; The Royal Society and Leverhulme Trust, United Kingdom; Eric and Wendy Schmidt Fund for Strategic Innovation, United States of America.

In addition, individual members wish to acknowledge support from Chile: Agencia Nacional de Investigaci\'on y Desarrollo (ANID FONDECYT reg. 1230987, FONDECYT 1230812, FONDECYT 1240864, Fondecyt 3240661, Fondecyt Regular 1240721); China: Fundamental Research Funds for the Central Universities (010-63263105), Chinese Ministry of Science and Technology (MOST-2023YFA1605700, MOST-2023YFA1609300), National Natural Science Foundation of China (NSFC 12275265, NSFC-W2543005); Czech Republic: Czech Science Foundation (GACR - 24-11373S), Ministry of Education Youth and Sports (ERC-CZ-LL2327, FORTE CZ.02.01.01/00/22\_008/0004632); EU: H2020 European Research Council (ERC - 101002463); European Union: European Research Council (BARD No. 101116429, ERC - 101219398, ERC 101089007), European Regional Development Fund (HE COFUND GA No.101081355, ERDF), Marie Sklodowska-Curie Actions (GAP-101168829); France: Agence Nationale de la Recherche (ANR-22-EDIR-0002, ANR-24-CE31-0504-01); Germany: Deutsche Forschungsgemeinschaft (DFG - 469666862); China: Research Grants Council (GRF); Italy: Istituto Nazionale di Fisica Nucleare (LHC-MIUR - 28003/2025), Ministero dell'Università e della Ricerca (NextGenEU I53D23000820006 M4C2.1.1, SOE2024\_0000023); Japan: Japan Society for the Promotion of Science (JSPS KAKENHI  JP25H0063, JSPS KAKENHI JP22H04944, JSPS KAKENHI JP24K23939, JSPS KAKENHI JP24KK0251, JSPS KAKENHI JP25H00650, JSPS KAKENHI JP25H01291, JSPS KAKENHI JP25K01011, JSPS KAKENHI JP25K01023, JSPS KAKENHI JP25KK0047); Poland: Polish National Science Centre (NCN 2021/42/E/ST2/00350, NCN OPUS 2023/51/B/ST2/02507, NCN OPUS nr 2022/47/B/ST2/03059, NCN UMO-2019/34/E/ST2/00393, UMO-2022/47/O/ST2/00148, UMO-2023/49/B/ST2/04085, UMO-2023/51/B/ST2/00920, UMO-2024/53/N/ST2/00869); Spain: Agència de Gestió d'Ajuts Universitaris i de Recerca. (AGAUR - 2023 BP 00141), Ministry of Science and Innovation (RYC2019-028510-I, RYC2020-030254-I, RYC2021-031273-I, RYC2022-038164-I), Ministerio de Ciencia, Innovación y Universidades/Agencia Estatal de Investigaci\'on (EU NextGenerationEU (PRTR-C17.I1), PID2022-142604OB-C22); Sweden: Carl Trygger Foundation (Carl Trygger Foundation CTS 22:2312), Swedish Research Council (Swedish Research Council 2023-04654, VR 2021-03651, VR 2022-03845, VR 2022-04683, VR 2023-03403, VR 2024-05451, VR 2025-05940), Knut and Alice Wallenberg Foundation (KAW 2023.0366); United Kingdom: The Binks Trust, Royal Society (NIF-R1-231091); United States of America: U.S. Department of Energy (ECA DE-AC02-76SF00515), John Templeton Foundation (John Templeton Foundation 63206), Neubauer Family Foundation.

%
%


%
\clearpage
\appendix
\printbibliography
\clearpage
 
\begin{flushleft}
\hypersetup{urlcolor=black}
{\Large The ATLAS Collaboration}

\bigskip

\AtlasOrcid[0000-0002-6665-4934]{G.~Aad}$^\textrm{\scriptsize 102}$,
\AtlasOrcid[0000-0001-7616-1554]{E.~Aakvaag}$^\textrm{\scriptsize 17}$,
\AtlasOrcid[0000-0002-5888-2734]{B.~Abbott}$^\textrm{\scriptsize 121}$,
\AtlasOrcid[0000-0002-0287-5869]{S.~Abdelhameed}$^\textrm{\scriptsize 83b}$,
\AtlasOrcid[0000-0002-1002-1652]{K.~Abeling}$^\textrm{\scriptsize 54}$,
\AtlasOrcid[0000-0001-5763-2760]{N.J.~Abicht}$^\textrm{\scriptsize 48}$,
\AtlasOrcid[0000-0002-8496-9294]{S.H.~Abidi}$^\textrm{\scriptsize 30}$,
\AtlasOrcid[0009-0003-6578-220X]{M.~Aboelela}$^\textrm{\scriptsize 44}$,
\AtlasOrcid[0000-0002-9987-2292]{A.~Aboulhorma}$^\textrm{\scriptsize 36e}$,
\AtlasOrcid[0000-0001-5329-6640]{H.~Abramowicz}$^\textrm{\scriptsize 154}$
% ...
\bigskip
\\

% ...
$^{17}$Department for Physics and Technology, University of Bergen, Bergen; Norway.\\
% ...
$^{30}$Physics Department, Brookhaven National Laboratory, Upton NY; United States of America.\\
% ...
$^{36}$$^{(a)}$Facult\'e des Sciences Ain Chock, Universit\'e Hassan II de Casablanca;$^{(b)}$Facult\'{e} des Sciences, Universit\'{e} Ibn-Tofail, K\'{e}nitra;$^{(c)}$Facult\'e des Sciences Semlalia, Universit\'e Cadi Ayyad, LPHEA-Marrakech;$^{(d)}$LPMR, Facult\'e des Sciences, Universit\'e Mohamed Premier, Oujda;$^{(e)}$Facult\'e des sciences, Universit\'e Mohammed V, Rabat;$^{(f)}$Institute of Applied Physics, Mohammed VI Polytechnic University, Ben Guerir; Morocco.\\
% ...
$^{44}$Physics Department, Southern Methodist University, Dallas TX; United States of America.\\
% ...
$^{48}$Fakult\"{a}t Physik, Technische Universit{\"a}t Dortmund, Dortmund; Germany.\\
% ...
$^{54}$II. Physikalisches Institut, Georg-August-Universit\"{a}t G\"ottingen, G\"ottingen; Germany.\\
% ...
$^{83}$$^{(a)}$Khalifa University of Science and Technology, Abu Dhabi;$^{(b)}$New York University Abu Dhabi, Abu Dhabi;$^{(c)}$United Arab Emirates University, Al Ain;$^{(d)}$University of Sharjah, Sharjah; United Arab Emirates.\\
% ...
$^{102}$CPPM, Aix-Marseille Universit\'e, CNRS/IN2P3, Marseille; France.\\
% ...
$^{121}$Homer L. Dodge Department of Physics and Astronomy, University of Oklahoma, Norman OK; United States of America.\\
% ...
$^{154}$Raymond and Beverly Sackler School of Physics and Astronomy, Tel Aviv University, Tel Aviv; Israel.\\
% ...
$^{a}$ Also at Affiliated with an institute formerly covered by a cooperation agreement with CERN.\\
$^{b}$ Also at An-Najah National University, Nablus; Palestine.\\
$^{c}$ Also at Borough of Manhattan Community College, City University of New York, New York NY; United States of America.\\
$^{d}$ Also at Center for Interdisciplinary Research and Innovation (CIRI-AUTH), Thessaloniki; Greece.\\
$^{e}$ Also at Centre of Physics of the Universities of Minho and Porto (CF-UM-UP); Portugal.\\
$^{f}$ Also at CERN, Geneva; Switzerland.\\
% ...

\end{flushleft}


%

%
%

%

%
%
%
%

\end{document}